\documentclass[12pt]{jetp}
\begin{document}
\English


\title{Shadows from spinning black holes in extended gravity}

\setaffiliation1{Sternberg Astronomical Institute, Lomonosov Moscow State University, University Prospekt, 13, Moscow, 119234, Russia}

\setaffiliation2{Department of Quantum Theory and High Energy Physics, Physics Faculty, Lomonosov Moscow State University, Vorobievi Gory, 1/2, Moscow, 119234, Russia}

\setaffiliation3{Department of Astrophysics and Stellar Astronomy, Physics Faculty, Lomonosov Moscow State University, Vorobievi Gory, 1/2, Moscow, 119234, Russia}

\setauthor{S.~O.}{Alexeyev}{12}

\setauthor{V.~A.}{Prokopov}{13}

\abstract{

When the the shadow image of a supermassive black hole in the center of M87 was obtained a new era in the observational astrophysics began. The resolution improvement would allow to test extended gravity models at event horizon scales. We develop the method allowing to take into account the black hole rotation   in black hole shadow modelling. It is demonstrated that the shadow of a black hole with a diameter smaller than $4R_{Sw}$ ($R_{Sw} = 2M$) cannot be represented in the terms of pure Kerr-Newman metric. Therefore, the detection of a shadow of such size would indicate that further expansions must be taken into account. This fact could give new indications on the detailed form of the theory of gravity at the considered scales.}

\maketitle

\section{Introduction}

The general theory of relativity (GR) is a generally accepted theory of gravity and provides a correct description of the widest range of gravitational phenomena \cite{Will:2014kxa,TheLIGOScientific:2016src}. The existence of such phenomena as dark matter and dark energy testifies that GR correctly describes gravitational phenomena not in the all range of space-time scales \cite{Weinberg:1988cp,Clowe:2006eq,Ade:2015xua}. Therefore,  the basic physical principles for the modification of GR are widely discussed in the literature \cite{Capozziello:2011et,Berti:2015itd}. The spectrum of such extended theories consists from various models of $f(R)$ gravity, scalar-tensor theories, including Horndeski models \cite{Sotiriou:2008rp,DeFelice:2010aj,Charmousis:2011bf} and so on.

The model of extended gravity should be considered in the case of its better description of gravitational phenomena, at least at some space-time scale. Models that pass solar system tests are of particular interest. Namely, there are the models whose post-Newtonian (as well as post-Keplerian) parameters are consistent with empirical data \cite{Will:2014kxa,Berti:2015itd,Dyadina:2018ryl}. 

The first images of the supermassive black hole (BH) shadow in the center of the M87 galaxy \cite{Akiyama:2019cqa} open new possibility to test extended theories of gravity at the scales of event horizons where there is hope to extract the corrections to GR, rapidly decreasing with distance \cite{Zakharov:2014lqa,Zakharov:2018awx}.

The Schwarzschild metric (as the simplest model) can be extended to take into account the tidal charge, appearing as a contribution of additional dimensions\cite{Dadhich:2000am}. The Reissner-Nordstrom metric extended by the tidal charge can describe more BH configurations, since the sign of the tidal charge is opposite to the electromagnetic one\cite{Alexeyev:2010zzb}. Such extension represents the addition of the next expansion term against $1/r$ to the metric function. Generically the component $g_{00}$ can be performed in the form of infinite series against $1/r$. In fact, there are the Taylor ones, where the numerical coefficients are defined by the theory under consideration. For example, in the effective string gravity model with second-order curvature corrections (the Gauss-Bonnet model\cite{Alexeev:1996vs,Alexeyev:2012zz}) the numerical solution is approximated by such a polynomial. 

Earlier, in Ref. \cite{Alexeyev:2019keq} we investigated the consequences of a shadow model extension for the non-rotating BH. Now we generalize the model by including of the BH rotation. It is necessary to note that the rotation {\it distorts} the shape of the shadow, dividing the photon sphere into a set of different orbits. So each point on the edge becomes a unique probe of the BH potential. Therefore, it may become possible to determine or constraint the BH metrics parameters by the shape of the shadow.

In paragraph \ref{s2} we discuss the generating of a rotating solution from the non-rotating one, paragraph \ref{s3} is devoted to a shadow for a spinning BH, in paragraph \ref{s4} we discuss different possible forms of shadows, and the section \ref{s5} contains our conclusions.

All calculations are presented in the Planck units $G=c=\hbar=1$.

\section{Spinning Black Hole: Metric}\label{s2}

To obtain the metric of a rotating BH, we use the method proposed in \cite{Tsukamoto:2017fxq} and developed in \cite{SAVP}. With the help of the Newman-Janis algorithm\cite{1965JMP.....6..915N} we can transform the metric
\begin{eqnarray}\label{f1}
ds^2 = & - & \left( 1-\cfrac{2m(r)}{r}\right) dt^2 + \left( 1-\cfrac{2m(r)}{r}\right)^{-1} dr^2 \nonumber \\ & + & r^2 (d\theta^2+ \sin^2\theta d\phi^2 )
\end{eqnarray}
to the following one:
\begin{eqnarray}\label{f2}
ds^2 = & - & \left( 1-\cfrac{2m(r)r}{\rho^2}\right) dt^2 - \cfrac{4m(r)ar \sin^2\theta}{\rho^2} d\phi dt \nonumber \\ & + & \cfrac{\rho^2}{\Delta} dr^2 + \rho^2 d\theta^2 \nonumber \\ & + & \left(r^2+a^2+\cfrac{2m(r)a^2r\sin^2\theta}{\rho^2}\right)\sin^2\theta d\phi^2 , 
\end{eqnarray}
where:
\begin{eqnarray}\label{f3}
&& \rho^2=r^2+a^2\cos^2\theta , \nonumber \\
&&  \Delta(r)=r^2-2m(r)r+a^2 ,
\end{eqnarray}
$a=J/M$ is the spin of a BH, $M$ is its mass and $J$ is its angular momentum. For
\begin{equation}\label{f4}
    m(r)=M-\cfrac{Q^2}{2r} ,
\end{equation}
where $Q$ is the electric charge. The discussed algorithm transforms the Reissner-Nordstrom metric into the Kerr-Newman one \cite{1965JMP.....6..915N,Newman:1965my}. Therefore for the BH with the tidal charge one obtains that
\begin{equation}\label{f5}
    m(r)=M-\cfrac{q}{2r} ,
\end{equation}
where $q$ is the tidal charge. In opposite to the Reissner-Nordstrom case $q$ can be positive or negative.

Further, it is important to note that in extended gravity model the Schwarzschild, Reissner-Nordstrom and Kerr-Newman metrics do not always serve as exact solutions playing the role of the leading order first approximation only. So, in the non-rotating case, when the metric is established as infinite Taylor series against $1/r$ one can write:
\begin{align}\label{f7}
  g_{00} = 1 - \cfrac{C_1}{r}+\cfrac{C_2}{r^2}+\cfrac{C_3}{r^3}+... ,
\end{align}
where $C_1$ is double BH mass, $C_2$  is its charge(s). If $C_2$ is not vanish its contribution decreases with distance faster than $C_1$ one. Such an effect appears to be significant in the strong gravitational lensing and BH shadows. The first image of BH shadow is already obtained \cite{Akiyama:2019cqa} but the resolution is not enough to analyze the shape of the shadow. Therefore, one requires in increased observational accuracy to extract additional corrections.

If one applies the Newman-Janis algorithm to Eq. (\ref{f7}), the resulting metric (\ref{f2}) would contain the following term:
\begin{align}\label{f8}
   m(r)=M-\cfrac{q}{2r}-\cfrac{C_3}{2r^2}-...-\cfrac{C_n}{2r^{n-1}}-... .
\end{align}
We limit our consideration by $C_3$.

\section{The shadow of the rotating black hole}\label{s3}

According to the Ref. \cite{Tsukamoto:2017fxq} for constructing of the shadow of a rotating BH one needs in the following functions:
\begin{equation}\label{f9}
\begin{gathered}
     \xi_- = \cfrac{4m_0 r^2_0 - (r_0 +f_0 m_0 )(r^2_0 +a^2_0) )}{a(r_0 - f_0 m_0 )} ,
     \\
    \eta_- = \cfrac{r^3_0 [4(2-f_0 )a^2 m_0 -r_0(r_0 - (4-f_0 )m_0 )^2]}{a^2(r_0 - f_0 m_0 )^2} ,
  \end{gathered}
\end{equation}
where
\begin{equation}\label{10}
     f_0 = 1+\cfrac{m'(r_0)r_0}{m(r_0)} ,
\end{equation}
$r_0$ represents the the photon orbit radius corresponding to the shadow edge. For each point the value of $r_0$ is unique therefore the coordinates $[\alpha,\beta]$ on the image plane looks like \cite{Tsukamoto:2017fxq}: 
\begin{equation}\label{f11}
\begin{gathered}
     \alpha= \cfrac{\xi_-}{\sin{\theta_i}} , \\
    \beta = \pm \sqrt{\eta_- +(a-\xi_-)^2-\left(a\sin{\theta_i} - \cfrac{\xi_-}{\sin{\theta_i}}\right)^2} ,
  \end{gathered}
\end{equation}
where $\theta_i$ represents the angle between the rotational axis and the line of sight. The sign `` + '' corresponds to the upper part of the shadow, and the sign `` - '' corresponds to the lower one. 

\begin{figure}[!hbtp]
\begin{center}
\includegraphics[width=\linewidth]{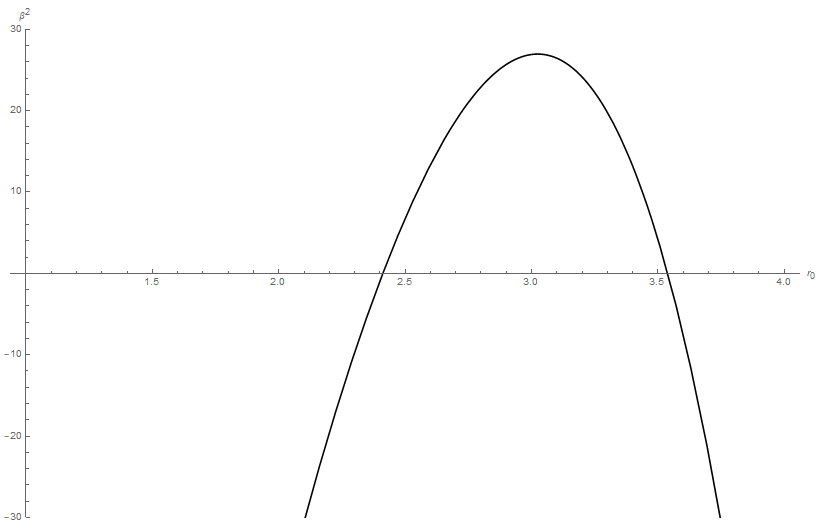}
\end{center}
\caption{The range of the value $r_0$ where $\beta^2 > 0$.}
\label{p1}
\end{figure}

To obtain the range of $r_0$ corresponding to shadow edge it is necessary to solve the following equation:
\begin{equation}\label{f12}
    \beta^2 =  0,
\end{equation}
and to take two maximal roots with $\beta^2>0$ (see Fig.\ref{p1}). We did this numerically.

So, for the metric (\ref{f2}) with (\ref{f8}) the functions $\xi_-$ and $\eta_-$ take the following form:
\begin{eqnarray}\label{f13}
\xi_- & = & \cfrac{4r^2_0 \xi_A - (r_0^2 + a^2) \xi_B}{a \xi_C} , \nonumber \\
\eta_- & = & \cfrac{r_0^3[\eta_A a^2 - r_0 \eta_B^2]}{a^2 \eta_C^2} ,
  \end{eqnarray}
where
\begin{eqnarray*}
\xi_A & = & M - \cfrac{q}{2r_0} - \cfrac{C_3}{2r_0^2} - \ldots - \cfrac{C_n}{2r_0^{n-1}} - \ldots ,\\
\xi_B & = & r_0 + M + \cfrac{C_3}{2r_0^2} + \ldots + \cfrac{(n-2)C_n}{2r_0^{n-1}} + \ldots ,\\ 
\xi_C & = & r_0 - M - \cfrac{C_3}{2r_0^2} - \ldots - \cfrac{(n-2)C_n}{2r_0^{n-1}} - \ldots , \\
\eta_A & = & 4M - \cfrac{4q}{r_0} - \cfrac{6C_3}{r_0^2} - \ldots - \cfrac{2n C_n}{r_0^{n-1}} - \ldots , \\
\eta_B & = & r_0 - 3M + \cfrac{2q}{r_0} + \cfrac{5 C_3}{2r_0^2} + \ldots \\ && + \cfrac{(n+2)C_n}{2r_0^{n-1}} + \ldots , \\
\eta_C & = & r_0 - M - \cfrac{C_3}{2r_0^2} - \ldots - \cfrac{(n-2)C_n}{2r_0^{n-1}} - \ldots ,
\end{eqnarray*}
To simplify the model we normalize the metric by the BH mass $M$. There the values of $r$ and $a$ are measured in the units of $M$ and $C_n$ are measured in the units of $M^n$.

\section{Shadow analyze}\label{s4}

\begin{figure}[!hbtp]
\begin{center}
\includegraphics[width=\linewidth]{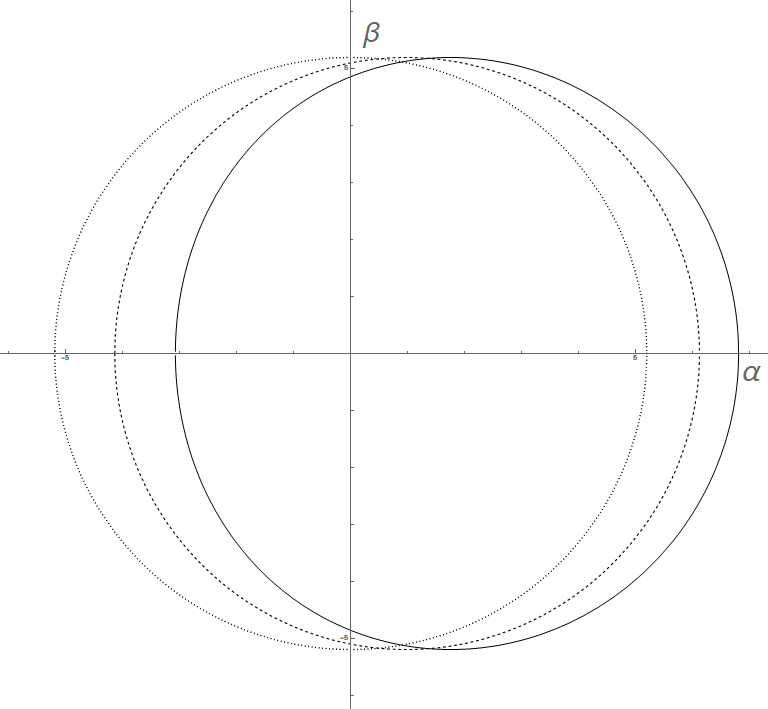}
\end{center}
\caption{The dependence of the shape of the shadows of a black hole during the rotation. The inclination angle is equal to $\theta_i=-\pi/2$, the charge and the third correction are $q=0.17$, $C_3=-0.5$ correspondingly. Dotted line represents the non-rotating case ($a=0$), dashed one corresponds to a medium rotating case $a=0.5$, solid one is a fast rotation case $a=0.9$. The rotation axis is directed along $\beta$.}
\label{p2}
\end{figure}

\begin{figure}[!hbtp]
\begin{center}
\includegraphics[width=\linewidth]{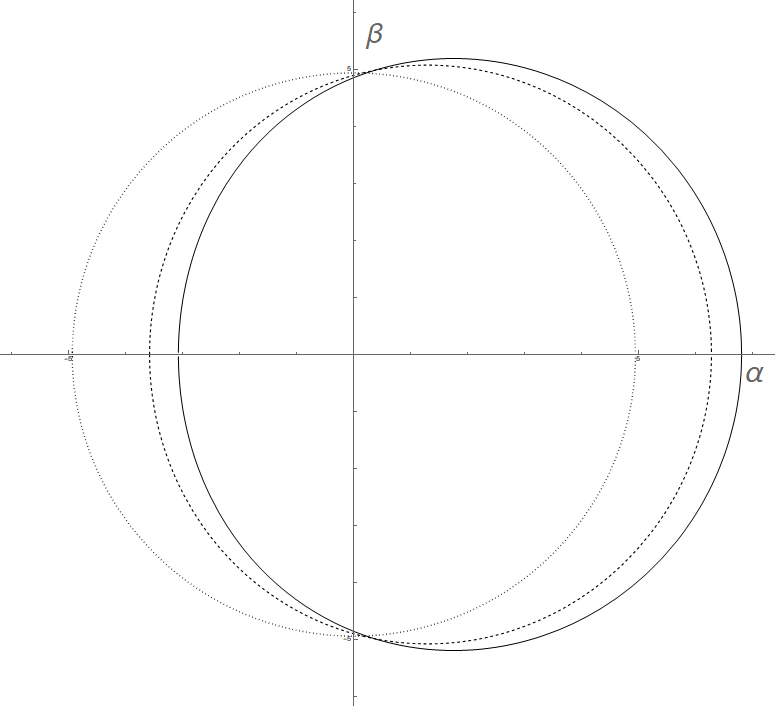}
\end{center}
\caption{The dependence of the shape of the shadows of a black hole on the inclination angle. Charge, third correction and spin of the black hole are  $q=0.17$, $C_3=-0.5$, $a=0.9$. Dotted line corresponds to the inclination angle $\theta_i=0$, dashed one corresponds to the inclination angle $\theta_i=-\pi/4$, solid one corresponds to the inclination angle $\theta_i=-\pi/2$. The axis of rotation is directed along $\beta$.}
\label{p3}
\end{figure}

Here it is necessary to note that we define the ``size of a shadow'' as its diameter along the axis of rotation. If the inclination angle is equal to $\theta_i = \pm \pi / 2$ the shadow size in the rotating case becomes the same as for the non-rotating one (Fig.\ref{p2}). When this angle decreases the size of the shadow decreases also (Fig.\ref{p3}). In the vanishing inclination ($\theta_i = 0$) case the shadow becomes circularly symmetrical (like in the non-rotating case) but slightly compressed due to the rotation. Previously \cite{Alexeyev:2019keq} we demonstrated that two non-rotating BHs with different parameters could have the same shadow size. In the rotating case each point of the shadow edge can be treated as the image of the BH's potential since it corresponds to the unique photon orbit radius $r_0$. If $r_0$ range is growing, it becomes easier to find differences in BH potentials. That is why the angle of inclination value equal to $\theta_i = \pm \pi / 2$ provides the best viewpoint for the distant observer. In the case $\theta_i = 0$ it is impossible to distinguish the shadow types due to charge, spin, or extension parameters without any additional observations.

\begin{figure}[!hbtp]
\begin{center}
\includegraphics[width=\linewidth]{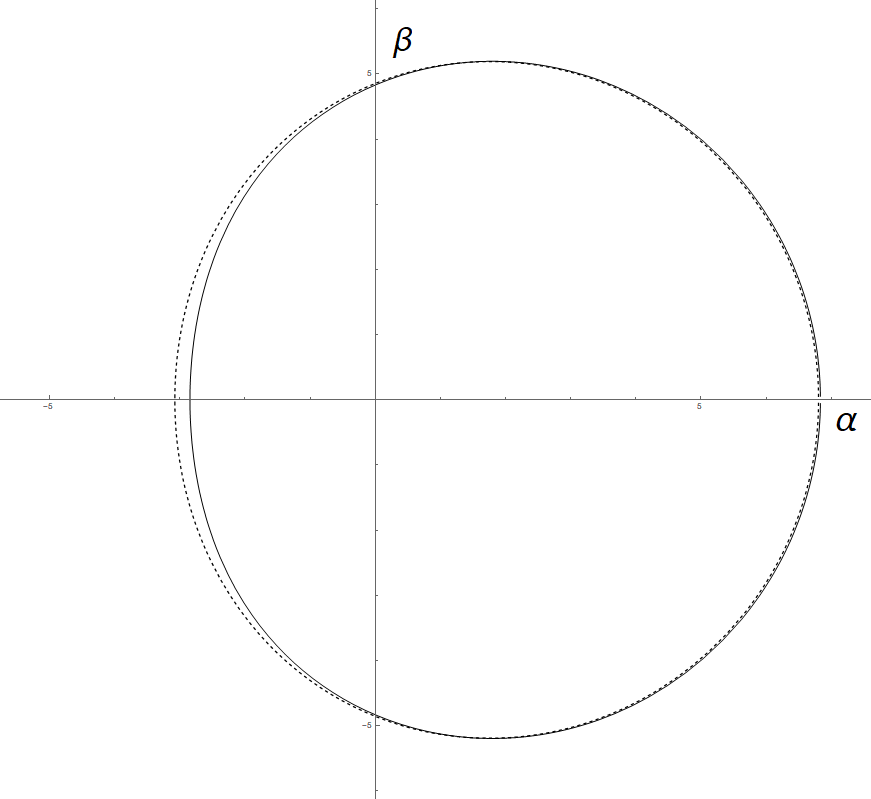}
\end{center}
\caption{Black holes with the spin equal to $a=0.9$ and the following parameters: dashed line corresponds to $q=0.17$, $C_3=-0.5$, solid one corresponds to $q=0$, $C_3=0$. The rotation axis is directed along the $\beta$ one, the inclination angle is equal to $\theta_i=-\pi/2$.}
\label{p4}
\end{figure}

It is interesting to compare the shadow models of BH at $\theta_i = - \pi / 2$ with different other parameters. For example, Kerr BH ($q = 0$ and $C_3 = 0$) and spinning one with $q = 0.17$ and $C_3 = -0.5$. Note that in the absence of a rotation their shadows are the same. The distance between the two corresponding points of the edges increases while the spinning moment becomes greater. At the spin value equal to $a = 0.5$ the maximum distance tends to $0.05M$, which represents about $0.5\% $ of a shadow size. For a strongly rotating BH with $a = 0.9$ (Fig. \ref{p4}) the maximum distance tends to $0.2M$ ($2\% $ of the shadow size, about $1 \mu as $ for Sag A* and M87*). Therefore BHs with larger spin are better for extended gravity tests. 

\begin{figure}[!hbtp]
\begin{center}
\includegraphics[width=\linewidth]{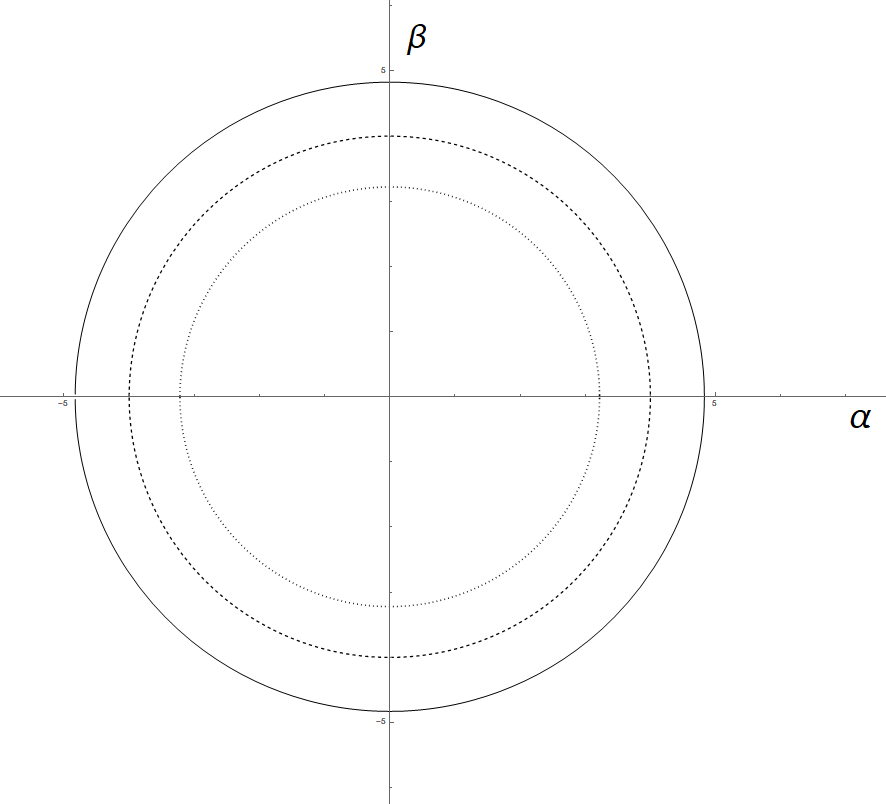}
\end{center}
\caption{The minimal sizes of the spinning BH shadow and different values of the tidal charge, visible at the angle of inclination equal to $\theta_i=0$. Solid line corresponds to the minimum shadow size of the Kerr black hole ($q=0$, $C_3=0$, $a=1$). Dashed line corresponds to the minimum Kerr-Newman black hole size ($q=1$, $C_3=0$, $a=0$). Dotted line corresponds to a black hole whose metric includes the third correction($q=1.4$, $C_3=-0.5$, $a=0.5$).}
\label{p5}
\end{figure}

In opposite the worst position is: $\theta_i = 0$. However, in the Kerr-Newman case the minimal BH shadow size (we use ``naked singularities'' as the limit in our consideration) is reached at $q = 1$, $a=0$ which corresponds to $4R_{Sw}$ (Fig. \ref{p5}). As we demonstrated earlier\cite{Alexeyev:2019keq} in minimally extended BH model the shadow sizes less than $4R_{Sw}$ appear. Such BH does not exist in Kerr-Newman model and could be the indication of extended gravity appearance.

\section{Conclusions}\label{s5}

The resolution achieved by the Event Horizon Telescope in the M87 observation is equal to about half of the shadow size ($20\mu as$), which is still not enough to measure the BH spin \cite{Akiyama:2019cqa}. The improving of the resolution at further observations would allow to use them for testing and/or selecting of extended gravity models, probably becoming some kind of PPN formalism \cite{Will:2014kxa}. However, the unique cases could realize when the shadow size would be less than $4R_Sw$ and for the shadow modeling it seems to be not enough to select between other possible reasons (for example, \cite{Perlick:2015vta,Bisnovatyi-Kogan:2018vxl}). Such BH shadows would show the importance of extended gravity theories.

\section*{Acknowledgments}

The authors acknowledge the support from the Program of development of M.V. Lomonosov Moscow State University (Leading Scientific School `Physics of stars, relativistic objects and galaxies').

\end{document}